\documentclass{rspublic}

\begin{document}

\title{Magnetic Activity in Stars, Discs \& Quasars}

\author{D. Lynden-Bell}

\affiliation{Institute of Astronomy, Cambridge CB3 0HA \& Clare
College\\ PPARC Senior Fellow seconded to the Physics Dept., The
Queens University, Belfast}

\label{first page}

\maketitle

\begin{abstract}{MAGNETISM. JETS, $\gamma $-RAY BURSTS, CORONAL HEATING.}

\end{abstract}

\section{Introduction}

Although magnetic fields in interstellar matter were postulated almost
fifty years ago, magnetohydrodynamic theory was then much hampered by
our inability to see what the magnetic field configurations were like
and, after a decade of innovative development, cynics, not without
some justification, began to claim that {\bf anything} can happen when
magnetism and an imaginative theorist get together.  Thus cosmic
lightning in particular received a bad press.  More recently great
advances in observational techniques that we shall hear of from Title,
Beck, Moran and Mirabel have enabled us to see not only the sun's
magnetic field with unprecedented clarity but the fields in galaxies,
quasars and microquasars are now measured and not merely figments of
fertile imaginations.  Let me return to the beginnings of the subject.

\section{History}

Passing by the earthly endeavours of Gilbert, Michell, Gauss and
others in earlier centuries our subject sprang to life with Hale's
(1914) discoveries of kilogauss magnetic fields in sunspots and his
laws governing their polarities in successive sunspot cycles.  His
invention of the spectroheliograph also gave the first pictures of the
magnetically dominated chromospheric structures.  After a long struggle
the Babcock's (1955) finally developed a magnetograph sensitive
enough to measure the `general' field of the sun.  It was commonly
believed to be there because the polar plumes seen in eclipse
photographs at sunspot minimum look remarkably similar to iron filings
around a bar magnet.  However in 1961 H.W. Babcock went on to show that
the sun's general field reversed every 11 years - far far shorter than
the many magennia required for the diffusion of a magnetic field
through a solid conductor.  Babcock's paper, which should be read by
all those interested in solar magnetism, went on to explain how
reconnections in the solar atmosphere can lead to flux expulsion and
he modelled the topology of the field throughout the 22 year cycle.

\section{The Heating of the Corona}

Spectroscopy of the Corona during total eclipses has a long history
too.  My wife owes her life to it since her grandparents first met
when expeditions from Harvard and Cambridge, England, set up their
1905 eclipse apparatus in the same Spanish field!  In 1931 Grotrian
derived a million degree temperature for the Corona.  However it
took until the second world war before the high temperatures of the
Chromosphere and the Corona were agreed.  By then the pioneering work
of Grote Reber (1944) who built his own radio telescope, had also
shown million degree temperatures.  How can it be that the outer
atmosphere of the sun is so much hotter than the photosphere although
the heat comes from inside?  Fred Hoyle (1959) suggested it was the
splash of the interstellar gas falling onto the sun but soon afterward
the solar wind was found to blow in the opposite direction.  Lighthill
had developed the theory of sound generated by turbulence and it was
thought that non-linear sound waves generated in the convection zone
would build up into shocks and so might heat the corona by mechanical
dissipation.  However this proved inadequate by a factor of thirty or
more, so it was soon abandoned in favour of a more sophisticated theory
of magnetohydrodynamic waves.  Here the magnetic field acts like a
duct bringing the wave energy to greater heights and dissipation occurs
primarily at resonances.  Although such heating may still be
inadequate such wave motions have certainly been seen and this
mechanism has had quite a wide following in the solar physics
community.

However, Babcock's reversal of the whole flux of solar poloidal field
clearly required much dissipation as flux is reconnected at neutral
points or rather the critical lines that join them.  Thus in 1974
R.H. Levine wrote his paper `A New Theory of Coronal Heating'.
In this theory flares are only the largest and most rapid
reconnections of magnetic field and the generation of mildly
suprathermal particles by reconnections goes on at all scales.  Thus
this year is in fact the 25th anniversary of the microflaring idea.
The paper formed part of a Harvard thesis supervised by David Layzer
who undoubtedly played a part in instigating the investigation.  There
may be something of an NIH (not invented here) complex in the slowness
of the solar community to adopt Levine's explanation but it was only
with the Yohkoh satellite's observations, reviewed here by
Harra-Meunion, that the dominant role of magnetism in the heating of
the corona became clear, and even then reconnection was not obvious as
the magnetohydrodynamic waves also needed magnetism to duct their
energy up into the corona.  It is up to the very high resolution of
the Trace Satellite (cf Title's talk and Parnell's analysis) to show
us whether there is sufficient microflaring to heat the corona.

\section{Jets and Accretion Discs}

Many of the other talks we shall hear involve Accretion Discs.  These
are formed:\\

\smallskip
\noindent
1. 	around the giant black holes postulated in
galactic nuclei as a theory of quasars (Lynden-Bell 1969, Bardeen
1970, Rees 1984).  They were first definitively found there by Moran
one of our speakers (Miyoshi, Moran et. al., 1995).  Indeed the concept
of such giant black holes and the way they would eventually be found
first arose in the remarkable work of the Reverend John Michell, a
fellow of this Society, in 1784.\\

\smallskip
\noindent
2.	In X-ray binary stars that transfer mass onto the compact
  partner (Prendergast \& Burbidge 1968, Pringle \& Rees 1972, Shakura
  \& Sunyaev 1973).\\

\smallskip
\noindent
3.	 Around proto-stars (L\"ust 1952, Pringle 1981). \\

\smallskip
\noindent
4.	 In the exotic SS 443-like objects in some supernova remnants
(Margon 1984).\\

\smallskip
\noindent
5.	In the micro-quasars that give superluminal apparent motions
within the Galaxy.  (Mirabel \& Rodriguez 1998).  We shall hear of
these from Mirabel who discovered the motions.\\

\smallskip
\noindent
6. As neutron stars tear each other apart during their coalescence
to form a stellar mass black hole.  Such events are thought to be the
origins of the famous $\gamma $-ray bursts of which we shall hear more from
Martin Rees.

\bigskip

Jets we first discovered in the giant elliptical galaxy M87 (Curtis
1918) then in the first quasar 3C273; then following Rees's 1971
suggestion that radio lobes must be continuously fed in radio
galaxies, Cygnus A (Hargrave \& Ryle 1974) and NGC 6251 (Baldwin
et. al., 1979) and in many other radio galaxies \& quasars such as
3C345, a quasar notable for its violent variability in the optical
and 3C279 notable for the high energy of its sporadic $\gamma$-ray emission
thought to be associated with the jet pointing in our direction.

Although the enigmatic Herbig-Haro objects were seen in emission
before 3C273 was discovered their association with jets from
star-forming accretion discs was only established later as infra-red
and millimeter wavelength observations enabled astronomers to
penetrate the swathes of dust in which stars are formed.  The jets of
HH212 seen at 2.12 $\mu m$ and of HH30 `filmed' by the Hubble Space
Telescope are good examples.

\bigskip

WHY DO FLAT ACCRETION DISCS FORM VERY NARROW HIGHLY COLLIMATED JETS
PERPENDICULAR TO THEIR PLANES?

\bigskip

Rees's first suggestion (1971) was that strong electromagnetic waves
generated by pulsars would push plasma aside and channel out of a
surrounding medium.  When the circular polarisation expected from such
an explanation was not seen Blandford and Rees (1974) gave their twin
exhaust model of squirting relativistic plasma.  Not convinced that
this would collimate stably enough, in 1978 I tried collimation by
relativistic vortices in a thick accretion disc but had difficulty in
getting the beams narrow enough, meanwhile Lovelace (1976) had
concentrated on magnetic models in which the collimation arises from
the direction of the magnetic field that permeates the system.  A new
and exciting role for magnetism was found by Blandford \& Znajek
(1977) who showed how flux through a black hole could withdraw spin
energy from it and give electric currents of electrons and positrons
travelling into the hole from opposite sides.  Later Blandford \&
Payne (1982) showed that, if the meridional magnetic field lines
splayed by more than 30$^\circ$ to the spin axis, then an accretion
disc would drive a centrifugal wind which could be collimated later by
the field.  Blandford will review for us quasar jets and their
magnetic fields.  There are many magnetohydrodynamic jet simulations
starting with different initial conditions.  Shibuta and Uchida (1985,
86) like Lovelace start with a large scale pervading magnetic field.
While this must surely be there in their star formation applications,
it is less clear how such an external weak field can make the
precessing jets of SS 433 or the microquasars.  I therefore favour
mechanisms that can produce the collimation from the disc itself.  In
groping towards such a theory I have found magnetic towers that grow
taller with every turn of the accretion disc (Lynden-Bell 1996) but
others including Pudritz have beautiful simulations that produce truly
dynamical jets (Ouyed et. al., 1997).  However let me not leave you
with the idea that everything is understood.  Many will tell you that
toroidal loops of field pinch, whereas I showed in the above paper
that their magnetic energy {\bf does not change} when the whole
magnetic field is expanded in the direction perpendicular to the jet
axis.  In a delightful swashbuckling samurai manner Okamoto (1999) has
criticised everyone in sight over their proposed collimation
mechanisms.  ``How are jets collimated?'' is still a live question.

\section{$\gamma$-ray Bursts}

No introduction should be without provocative questions so let me end
with another.

When a body with binding energy ${\scriptstyle \frac {1}{2}}GM^2/R$
forms in its own dynamical timescale $(GM/R^3)^{-{\scriptstyle 
\frac {1}{2}}}$ the rate of emission of the energy will be the ratio
${\scriptstyle \frac {1}{2}}G^{{\scriptstyle \frac
{3}{2}}}M^{{\scriptstyle \frac {5}{2}}}/R^{{\scriptstyle \frac {5}{2}}}$
which may be rewritten in terms of its final velocity dispersion $v$
where ${\scriptstyle \frac {1}{2}}Mv^2 = {\scriptstyle \frac {1}{
2}}GM^2/R$ by the Virial theorum.  Hence the rate of emission
${\dot E} = {\scriptstyle \frac {1}{2}}v^5/G$.
For $v=c$ this gives an emission rate of $1.8 \times 10^{57} ergs/s$
or $1.8 \ 10^{50}$ watts.  This is far greater than the emission rate
of even the most powerful $\gamma$-ray bursts; thus either ``we ain't seen
nothing yet'' or more likely $\gamma$-ray bursts are only radiating a tiny
fraction of the total energy involved the rest being swallowed into a
black  hole or emitted in neutrinos \& gravitational waves.  Certainly
the problems concerned with absorption by baryons are avoided if $\gamma$-ray
bursts come from the ends of long jets along which the energy flows as
electromagnetic Poynting flux thus converting only the magnetic field
energy.  

Do $\gamma$-ray bursts come from the heads of magnetically driven
relativistic jets formed by the accretion discs of binary neutron
stars coalescing to form a black hole?

Sir Martin Rees will be introducing us to these exotic events and
giving his latest ideas about them.

\end{document}